\documentclass{article}

\usepackage{arxiv}

\usepackage[utf8]{inputenc} % allow utf-8 input
\usepackage[T1]{fontenc}    % use 8-bit T1 fonts
\usepackage{url}            % simple URL typesetting
\usepackage{booktabs}       % professional-quality tables
\usepackage{amsfonts}       % blackboard math symbols
\usepackage{amsmath}
\usepackage{nicefrac}       % compact symbols for 1/2, etc.
\usepackage{microtype}      % microtypography
\usepackage{lipsum}		% Can be removed after putting your text content
\usepackage{graphicx}
\usepackage{natbib}
\usepackage{doi}
\usepackage[affil-it]{authblk}

\usepackage{graphicx,graphbox, wrapfig,lipsum}

\usepackage{hyperref}
\hypersetup{
            colorlinks, linktocpage=true, pdfstartpage=1, pdfstartview=FitV,%
    hypertexnames=true, pdfhighlight=/O,%
   linkcolor=red, citecolor=blue, %pagecolor=RoyalBlue,%
   }
   
   \hypersetup{
    colorlinks,
    linkcolor={red!50!black},
    citecolor={blue!50!black},
    urlcolor={blue!80!black}
}

\usepackage[bitstream-charter]{mathdesign}
\usepackage{tcolorbox}
\newtcolorbox{mybox}{colback=blue!5!white,colframe=blue!55!black}

\title{Les Houches lectures on Theoretical Ecology:\\ high-dimensional models and extreme events}

\author[1,*]{Ada Altieri}

\affil[1]{Laboratoire Matière et Systèmes Complexes (\emph{MSC}), Université Paris Cité, CNRS, 75013 \ Paris, \ France }
\affil[*]{Corresponding author: \textsc{ada.altieri@u-paris.fr}}

\begin{document}
\maketitle

\begin{abstract}
The study of ecological systems is gaining momentum in modern scientific research, driven by an abundance of empirical data and advancements in bioengineering techniques. However, a full understanding of their dynamical and thermodynamical properties, also in light of the ongoing biodiversity crisis, remains a formidable endeavor.
From a theoretical standpoint, modeling the interactions within these complex systems -- such as bacteria in microbial communities, plant-pollinator networks in forests, or starling murmurations -- presents a significant challenge. Given the characteristic high dimensionality of the datasets, alternative elegant approaches employ random matrix formalism and techniques from disordered systems. In these lectures, we will explore two cornerstone models in theoretical ecology: the MacArthur/Resource-Consumer model, and the Generalized Lotka-Volterra model, with a special focus on systems composed of a large number of interacting species. In the second part, we will highlight timely directions, particularly to bridge the gap with empirical observations and detect macroecological patterns.
\end{abstract}

\vspace{10pt}
\noindent\rule{\textwidth}{1pt}
\tableofcontents
\noindent\rule{\textwidth}{1pt}
\vspace{10pt}
%%%%%%%%%% END TODO: TOC

%%%%%%%%% TODO: CONTENTS 
% Write your article contents here, starting from first \section.
% An example structure is given below.

\section{Introduction}
\label{sec:intro}

In recent years, the study of ecological communities has attracted growing interest from both theoretical physics and quantitative biology. Despite significant advances, several foundational questions remain unresolved. Among these is the long-standing \emph{diversity-stability debate} \cite{mccann_diversitystability_2000, hatton_diversity_2024}, which seeks to clarify the role of ecological diversity in maintaining ecosystem resilience, once appropriate diversity metrics are defined. Another critical challenge lies in understanding how ecosystems respond to external perturbations, whether from sudden shocks or gradual environmental changes.
Ecosystems exhibiting abrupt responses to gradual environmental changes are particularly concerning. Prominent examples include desertification in arid regions, coral reef bleaching, and the collapse of tropical forests into treeless landscapes. These shifts are frequently linked to the emergence of multiple stable states under specific conditions \cite{van2021, amor2020, aguade2024taxonomy}. Minor perturbations can induce transitions between these states, triggering severe declines in biodiversity and ecosystem functionality. This raises a central theoretical question: Is the number of feasible equilibria in such ecosystems exponential or sub-exponential relative to system size?\footnote{In disordered systems literature, two main reference classes are commonly distinguished: i) spin-glass landscapes, characterized by a sub-exponential number of free-energy minima in the system size, with free-energy barriers expected to be sub-extensive; ii) simple structural-glass landscapes, where the number of free-energy minima grows exponentially with the system size, and the barriers separating these minima are extensive. This kind of question might also be rephrased and adapted to ecological systems.} Local multistability refers to the coexistence of numerous stable states, some of which may require an extended time to equilibrate. Whether the number of such stable and uninvadable states grows exponentially with the number of species -- a behavior that could be observed in simulations or tailored laboratory experiments -- or on a much smaller scale remains an open question.

Moreover, identifying key drivers of ecosystem fragmentation and segregation, as well as developing quantitative precursor indicators, could be essential for predicting ecological collapses. For instance, dispersal in spatially structured communities can rescue fragmented ecosystems from extinction. Depending on demographic fluctuations, the transition from a self-sustained state to ecosystem collapse may be continuous or abrupt, with the latter often associated with tipping points, metastability, and hysteresis -- concepts widely recognized not only in ecology but also in environmental science, and condensed matter physics.

In these lecture notes, we aim at bridging the gap between statistical physics approaches and the empirical analysis of large, heterogeneous ecosystems. 
Statistical physics offers efficient tools for investigating complex dynamics of species-rich ecosystems and eventually addressing still open questions in this booming field.
Notably, we shall discuss three reference models: neutral models, the Generalized Lokta-Volterra model, and the Consumer-Resource (or MacArthur) model. 

\begin{itemize}
\item \textbf{Neutral Models.}
Neutral models, exemplified by Hubbell's unified neutral theory of biodiversity and biogeography \cite{hubbell_unified_2011}, assume that all individuals are ecologically equivalent, regardless of species identity. They share identical per-capita birth and death rates, as well as equal migration and speciation probabilities. While this assumption implies that species are statistically indistinguishable, demographic stochasticity creates variability: species with larger populations are more likely to persist, while smaller populations face a higher risk of extinction due to random fluctuations. These dynamics are often modeled using stochastic birth-death processes, and in some cases, their stationary distributions can be exactly obtained. As such, neutral models capture fluctuations that are driven purely by demographic stochasticity.
Neutral theories have since been extended to incorporate additional factors and have successfully reproduced several macroecological patterns \cite{azaele_statistical_2016, chave_neutral_2004, rosindell_case_2012}.
However, the assumption of \emph{functional equivalence} among species has been widely criticized for oversimplifying biological systems. In fact, species often exhibit significant differences in their environmental adaptations, and their interactions actively shape the environment they inhabit.

\item \textbf{Niche models.} Conversely, models that explicitly account for species differences are collectively referred to as niche models.
Each niche is occupied by a single species according to the competitive exclusion principle.
In other words, two species competing for the same limited resource cannot coexist: the fittest one will dominate in the long term, thus leading either to extinction or an evolutionary shift towards another niche.
Among this class of models, Lotka-Volterra equations deserve a special place \cite{lotka1925, wangersky1978lotka, hofbauer_evolutionary_1998}. They were introduced in the early $20$th century --  in terms of a pair of first-order nonlinear differential equations --
postulating that the demographic rates of a species depend not only on its abundance but also on the abundance of other species in the community. This departure from neutrality enables niche models to explore how inter-specific interactions drive coexistence, competition, and overall stability of a given ecological community.

 However, one of the main shortcomings of the niche framework is its limited predictive ability. While it provides a way to explain the coexistence of two species by comparing their ecological requirements and preferences, it offers little insight into determining the number of niches in a given environment without directly counting the species that occupy them. MacArthur attempted to address this issue by proposing a famous consumer-resource model. Accordingly, a stable equilibrium can only be achieved if the number of competing species does not exceed the total available resources. However, this upper bound is frequently violated in natural ecosystems, particularly in planktonic communities, giving rise to the so-called \emph{paradox of the plankton} \cite{hutchinson1961paradox}.
 
\end{itemize}

By focusing on two of these benchmark models, we aim at achieving a better understanding of the intricate dynamics of large ecosystems, thus offering analytical insights into their resilience to external perturbations, species coexistence, and biodiversity maintenance.

\section{MacArthur's consumer-resource (CR) model: \\
a simple instance for niche theory}

The foundational concepts of competition and selection shaping communities can be traced back to the pioneering works of R. Levine and R. MacArthur in the sixties \cite{levine-macarthur1967, macarthur_species_1970}.

Let us consider $S$ species, characterized by their abundances $N_i$ ($i=1,2,...S)$ and competing for $M$ resources with abundances $R_\alpha$ ($\alpha=1,2...,M)$. Their dynamics are defined by:
\begin{equation}
\begin{split}
&\frac{d N_i}{dt}=N_i g_i(\textbf{R}) \ ,\\
& \frac{dR_\alpha}{dt}=F_\alpha (\textbf{R})-\sum_i N_i c_{i \alpha} R_\alpha=F_\alpha(\textbf{R})+\sum_{i}N_i q_{i \alpha}(\textbf{R}) \ ,
\end{split}
\end{equation}
where $c_{i \alpha}$ is a $S \times M$ matrix accounting for the metabolic strategies or consumer preferences in terms of the $\alpha$ resource, whereas $g_i(\textbf{R})=\sum_\alpha c_{i \alpha} w_\alpha R_\alpha - \chi_i$ is the growth rate  with quality factor $w_\alpha$. In the spirit of MacArthur's seminal work, each species has its own minimum requirement $\chi_i$. The function $F_\alpha(\textbf{R})$ is also called resource supply, which typically satisfies a logistic form with $F_\alpha(R_\alpha)=R_\alpha(K_\alpha-R_\alpha)$, $K_\alpha$ being the carrying capacity associated with resource $\alpha$ .
The formulation of such Consumer-Resource (CR) equations relies on the assumption that interactions between species are mediated by the consumption of external resources. In other words, the growth function and the impact vector for each species $q_{i \alpha}(\textbf{R})$ do not depend on the population size.

Given this system of $S \times M$ coupled equations, one can define three different conditions:
\begin{itemize}
\item{steady state population, by imposing $\frac{d N_i}{dt}=0$;}
\item{non-invasibility if $g_i(\textbf{R}) \le 0$;}
\item{feasible equilibria provided $N_i \ge 0$.}
\end{itemize}

\subsection{Low-dimensional criterion for feasibility and stability vs ZNGIs}

Pioneering works in theoretical ecology have relied on models composed of a few species and few resources. In these cases, stability properties could be deduced by simple geometrical arguments \cite{tilman_resources_1980, cui_houches_2024}. 
By denoting with $c_{i \alpha} R_\alpha=-q_{i \alpha}(\textbf{R})$ the \emph{impact vector} of species $i$, one can deduce fixed-point properties.
For the sake of simplicity, let us consider a $2$-dimensional model. There exist three geometric conditions that allow us to determine stable coexistence of two competing species:
\begin{itemize}
\item{The Zero Net Growth Isoclines (ZNGI), defined by the conditions $g_i(\textbf{R})=0$, must intersect. This criterion is valid in any dimension and leverages the \emph{Competitive Exclusion Principle} according to which at most $M$ consumer species can safely coexist.}
\item{The \emph{supply vector} $F_\alpha(\textbf{R})$ evaluated at the fixed point must lie within the cone spanned by the negative impact vector. This second condition guarantees the existence of a set of positive population sizes, with $N_i >0$. It corresponds to imposing $\frac{d R_\alpha}{dt}=0$, which leads to $F_\alpha(R_\alpha) = \sum_i N_i(-q_{i \alpha}(\textbf{R}))$, still holding regardless of the dimensionality of the system.}
\item{The impact of each species is biased toward the resource that significantly influences its growth rate, which is particularly challenging to identify in higher dimensions. Based on Tilman's geometrical framework, the orientation of the impact vector should be compared to that of the ZNGIs.}
\end{itemize}

\subsection{Rephrasing the CR model in high dimensions: statistical physics lens}

The aforementioned geometrical criteria offer an intuitive starting point for ecological interpretation, especially in the context of low-dimensional systems.
However, in recent years, there has been a growing focus on systems composed of an enormous number of interacting components, i.e. by considering $S, M \rightarrow \infty$ at fixed ratio $\gamma=M/S$ \cite{Tikhonov2017, Batista2021, altieri_constraint_2019}. Such a picture leads to the neglect of individual details and zooms out on the typical behaviors of large communities. 
In other words, since measuring how single-species parameters scale with the system size is particularly hard, one can therefore resort to a coarse-grained description by averaging over a large number of (functionally equivalent) microscopic degrees of freedom. \emph{Effective models} thus retain only a few fundamental features of the real system, using them to capture general properties, such as macroscopic observables or collective dynamics.

The difficulty of inferring the interactions in diverse ecosystems, such as for microbial communities, stems from several factors: the high dimensionality of microbiome datasets, the shortage of long time-series experiments, the time-dependent nature of microbial interactions, and environmental filtering effects due to abiotic factors such as resources, temperature, and pH. Another challenge involves disentangling noisy fluctuations from true ecological signals in empirical data.

An alternative and elegant way out was originally pioneered by Robert May leveraging Random Matrix Theory to model species interaction networks: each entry of the adjacency matrix is drawn from a given probability distribution. 
This approach drastically reduces the number of model parameters to be estimated, from $S^2$ to just a few \cite{Allesina2012}.
Within a statistical physics scenario, one typically splits the consumer preference matrix into a deterministic part and a fluctuating contribution, that is:
\begin{equation}
    \langle c_{i \alpha}\rangle=\frac{\mu_c}{M} \ ,  \hspace{0.5cm} \langle (c_{i \alpha}-\langle c_{i \alpha}\rangle)^ 2\rangle=\frac{\sigma_c^2}{M} \ .
\end{equation}
\begin{equation}
    \langle \chi_i \rangle=\chi \ , \hspace{0.3cm}\langle (\chi_i-\langle \chi_i\rangle)^2\rangle=\sigma_\chi^2
\end{equation}
\begin{equation}
    \langle K_\alpha \rangle=K \ , \hspace{0.3cm} \langle (K_\alpha-\langle K_\alpha\rangle)^2\rangle=\sigma_K^2.
\end{equation}
 This description is consistent with a \emph{generalist} picture, meaning that all species tend to grab a large number of resources. This is also consistent with the scaling of the metabolic strategy coefficients $c_{i \alpha}$: the outflow associated with each resource decreases with the amount of resources itself. This precise rescaling of the random coefficients will appear much clearer from a formal point of view in the following derivations. In other words, the scaling with $M$ of the first two moments guarantees a proper thermodynamic limit.

\subsection{Steady-state solutions}
To derive the mean-field self-consistent equations, we ask what happens if we add an additional species and resource to the pre-existing pool of $S$ species and $M$ resources. Intuitively, this is equivalent to exploring whether a new species can invade the ecosystem.

Formally, by denoting with $N_0$ the new species and $R_0$ the new resource, we aim to study correlations between resource and species entangled dynamics. However, because we always leverage a high-dimensional formalism in which $S, M \gg 1$, considering a system with $S+1$ species and $M+1$ resources does not sensitively affect the resulting thermodynamic picture. This addition will result in a slight perturbation of the original system, to be studied in terms of two local susceptibilities:
\begin{equation}
    \chi=\frac{1}{M} \sum_\alpha \frac{\partial R_\alpha^{*}}{\partial K_\alpha} \ , \hspace{0.6cm} \nu=\frac{1}{S} \sum_i \frac{\partial N_i^{*}}{\partial g_i}=-\frac{1}{S}\sum_i \frac{\partial N_i^{*}}{\partial \chi_i} \ .
    \label{suscep}
\end{equation}
The first expression (left) encodes the change of the mean resource abundance if we slightly perturb the supply of all external resources, by summing over $\alpha$. Conversely, on the right, the other susceptibility quantifies the change of the mean species abundance upon decreasing their fitness cost (or increasing the growth rate). 
The so-called susceptibilities $\chi$ and $\nu$ can be advantageously measured in experiments, for instance investigating growth and community structure by the addition of a given molecule, i.e. glucose.

In the following, in Sec. \ref{cavity}, we will see how to explicitly derive the steady-state abundances for a simpler model, by using the so-called \emph{cavity method}. For the sake of compactness, we show here the resulting expressions for $N_0^{*}$ and $R_0^{*}$, to be obtained by means of the Gaussian nature of the $c_{i \alpha}$, i.e. $c_{i \alpha}=\frac{\mu}{M}+\sigma_c \eta_{i \alpha}$ where $\langle \eta_{i \alpha}\rangle=0, \langle \eta_{i \alpha} \eta_{j \beta}\rangle=\frac{\delta_{ij} \delta_{\alpha \beta}}{M}$:

\begin{equation}
\begin{split}
& N_0^{*}=\text{max} \left[ 0, \frac{g+\sigma_g z_0}{\gamma \sigma_c^2 \chi}\right] \\
& R_0^{*}=\text{max} \left[0, \frac{K_\text{eff}+\sigma_K{_\text{eff}} \tilde{z}_0}{1-\gamma^{-1}\sigma_c^2 \nu}\right]
\end{split}
\label{cavity_MC}
\end{equation}
where we have denoted as $z_0, \tilde{z}_0$ two Gaussian variables with zero mean and unit variance. Eq. (\ref{cavity_MC}) will thus represent two truncated Gaussian distributions for the typical species and resource variables.
$K_\text{eff}(\textbf{N})=K_\alpha-\sum_i N_i c_{i \alpha}$ denotes instead an \emph{effective resource capacity}.

The above expressions give rise to a straightforward interpretation: by integrating out the quenched disorder formally encoded in the matrix $c_{i \alpha}$ and $K_\alpha$, we obtain an effective description where both the growth rates and the carrying capacities are normally distributed. Depending on the sign of the invasion growth rate, $g+\sigma_g z_0$, two options are possible:
\begin{itemize}
\item{If such a quantity is strictly positive, the species will survive with an abundance proportional to the invasion growth rate; }
\item{otherwise, it will not be able to invade, thus going extinct in a finite time.}
\end{itemize}
An analogous reasoning can be applied to $K_\text{eff}$: if the overall quantity $K_\text{eff}+\sigma_{K_{\text{eff}}} \tilde{z}_0$ turns out to be negative, the new resource will be depleted; otherwise, it will be proportional to the carrying capacity itself. 

In the previous expressions, we have assumed that the depletion and growth rates share the same matrix structure. However, we can also consider a slightly different variant:
\begin{equation}
\begin{split}
& \frac{dN_i}{dt} = N_i \left[\sum_\delta c_{i \delta} w_\delta R_\delta -\chi_i \right] \\
& \frac{d R_\alpha}{dt}=\frac{r_\alpha}{K_\alpha} R_\alpha (K_\alpha-R_\alpha)-\sum_j d_{j \alpha} N_j R_\alpha \ .
\end{split}
\label{MacArth_eqs}
\end{equation}
\vspace{0.25cm}

\begin{figure}[h]
    \centering
    \includegraphics[scale=1.1]{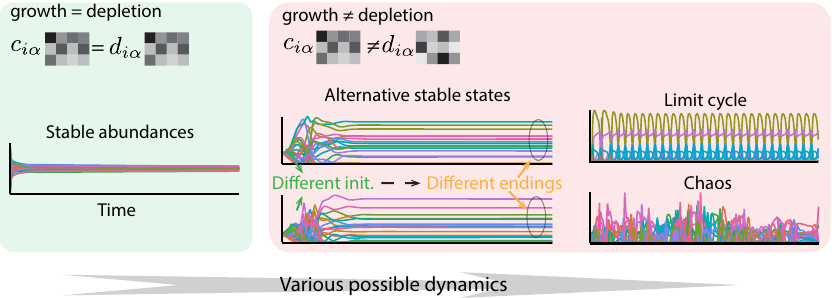}
    \caption{Distinct consumption and growth probabilities can lead to instabilities and subsequently
richer dynamics. Complex and out-of-equilibrium behavior, including both limit cycles and chaotic dynamics, can occur upon changing
the correlation between consumption and growth rates. Figure adapted from \cite{liu2025} (courtesy of Y. Liu).}
    \label{fig:MA_generalized}
\end{figure}
Imposing $\frac{dR_\alpha}{dt}=0$ leads to a stability criterion, i.e. $R_\alpha=K_\alpha -\frac{1}{m_\alpha} \sum_j d_{j \alpha} N_j $, where $m_\alpha$ denotes the ratio between the growth rate and the carrying capacity. The condition on $R_\alpha$ is associated with two distinct scenarios: i) if the term in parenthesis is negative, $\frac{dR_\alpha}{dt}<0$ for all positive $R_\alpha$, and therefore $R_\alpha=0$ will be the stable solution.
Conversely, if the term in parenthesis is positive, $R_\alpha=0$ becomes unstable and even a small perturbation will result in a positive growth rate. 
Despite its simplicity, the consumer-resource model can display a plethora of appealing dynamical behaviors.
Provided a feasible fixed point, it turns out to be stable if the two matrices are equal. In other words, by looking at the species abundance as a function of time, it will converge to a unique fixed point. 
More complex scenarios, like limit cycles and persistent fluctuations, can, however, emerge when the growth and depletion rates are not identical, as qualitatively depicted in Figure \ref{fig:MA_generalized}.

\subsubsection{Interpretation of the CR model as a Constraint Satisfaction Problem}

In the limit of a large number of species and resources $S, M \rightarrow \infty$, the MacArthur model has recently been reformulated as a constraint satisfaction problem, displaying the typical features of a perceptron model \cite{Tikhonov2017, altieri_constraint_2019}. Originally introduced in the theory of neural networks and machine learning as a binary classifier, the perceptron makes predictions by applying a weighted sum to the input features and using a threshold to assign a class label.
A key quantity is represented by the fractional volume of the solution space occupied by the interacting patterns, which corresponds to the maximum storage capacity of the network. In its convex regime -- meaning that the configurational landscape is characterized by a unique fixed point -- the critical transition occurs at zero interaction heterogeneity and fixed ratio between the number of patterns and the total degrees of freedom (i.e., the number of bits), equal to two. This transition separates a SAT region, where all constraints are simultaneously satisfied, from an UNSAT region, where no solution can fulfill all clauses.

Rephrased in an ecological context, the former of Eqs. (\ref{MacArth_eqs}) can be recast as a function of a resource surplus, $\Delta_i$:
\begin{equation}
\frac{dN_i}{dt} \propto N_i \Delta_i \ ,
\end{equation}
where again $i=1,...,S$.
This novel framework, introduced by Tikhonov and Monasson \cite{Tikhonov2017}, focuses exclusively on equilibrium configurations, leading to two possible scenarios: (i) $N_i > 0$ and $\Delta_i = 0$, or (ii) $N_i = 0$ and $\Delta_i < 0$ (extinction). Conversely, the opposite case, for $\Delta_i > 0$, is prohibited by the model definition.
Then, the total demand $T_\delta= \sum_i N_i c_{i \delta}$ -- where the index $\delta=1,...,M$ denotes the number of available resources -- is a function of the species abundances themselves, hence resulting in a feedback loop mechanism.

This model was initially devised to provide a clear geometric interpretation: metabolic strategies are represented as $S$ vectors in the $M$-dimensional space of resource availability. The equality $\sum_\delta c_{i \delta} w_\delta R_\delta = \chi_i$ defines a hyperplane, which divides the configuration space of abundances into two regions: an \emph{unsustainable} region, where the scalar product is smaller than the effective requirement $\chi_i$, and a \emph{sustainable} region, above this hyperplane, where a positive resource surplus enables species to self-sustain and multiply.
The single species requirement can also be recast in terms of a random cost, such that $
\chi_i=\sum_\delta c_{i \delta} + \epsilon x_i$, $\epsilon$ denoting an infinitesimally small parameter and $x_i$ a unit-variance Gaussian variable. Accordingly, one can establish the full phase diagram of the model and investigate its stability as a function
of a restricted number of control parameters: the ratio between the number of available resources and the total number of species $\gamma=M/S$; the width of the cost distribution $\epsilon$; the
average and variance of the metabolic strategy distribution; and, if an additional layer of complexity is introduced, the variance of the resource supply.
To characterize the system's dynamics, a Lyapunov function can be defined for each trajectory: 
\begin{equation}
\mathcal{F}(\lbrace N_i \rbrace)= \sum_\alpha R_\alpha \log \left( \sum_i N_i c_{i \alpha} \right) - \sum_i N_i \chi_i \ .
\end{equation}
One key advantage of defining a convex and bounded Lyapunov function is that it allows for the analysis of system stability in each phase by considering the harmonic fluctuations of such a function with respect to the species-abundance order parameter. In this system, it has been shown that a globally stable equilibrium state always exists, which, under certain conditions, may approach a marginally stable configuration \cite{altieri_constraint_2019}.

The asymptotic behavior of the system can then be evaluated in the large-$\beta$ limit, which corresponds to the zero-temperature regime of a statistical physics problem, where $\beta = 1/T$ is the inverse equilibrium temperature. 
As $\epsilon \rightarrow 0$ limit, the model exhibits a phase transition between two distinct regimes \cite{Tikhonov2017}: a shielded phase, where a collective and self-sustained behavior arises, basically unaffected by external conditions; and a vulnerable phase, where species cannot sustain themselves and turn out to be highly sensitive to environmental changes and improvements. This kind of transition reminds the aforementioned SAT/UNSAT transition occurring in the perceptron, interpreted as a linear signal classifier. 

Moreover, in the low-temperature regime, the logarithmic term in the effective free energy becomes the dominant contribution. This behavior is associated with vanishing $\Delta_i$ for a subset of selected species. 
Then, by analyzing the harmonic fluctuations of the Lyapunov function around its minimum, one can extract the spectral density in both phases, revealing a critical Marchenko-Pastur distribution \cite{mehta}.
This signals the onset of a marginal stability condition, reminiscent of the glassy phenomenology observed in critical jammed systems, such as infinite-dimensional hard spheres and other continuous constraint satisfaction problems belonging to the same universality class \cite{altieri_constraint_2019, altieri2019_jamming}.

\subsection{How to recover the Generalized Lotka-Volterra (GLV) model from a CR model?}

Different dynamical frameworks can be applied depending on the type of interaction being modeled—whether with the environment or among species. In the CR model, interactions primarily occur with the environment, whereas in the LV model, the environment serves as a mediator for inter-species interactions, akin to a thermal bath in physics.
For the sake of simplicity, let us consider the scenario in which both depletion and growth rates are encoded in the same matrix,  $c_{i \alpha}$.
\begin{equation}
\begin{split}
    &\frac{\dot{N}_i}{N_i}=\sum_\delta c_{i \delta} R_\delta - \chi_i \\
    & \frac{\dot{ R}_\alpha}{R_\alpha}=\frac{r_\alpha}{K_\alpha}(K_\alpha-R_\alpha)-\sum_j c_{\alpha j}^{T} N_j
  \end{split}  
\end{equation}
Each row of the matrix is an $M$-dimensional vector, the uptake profile of the species $i$. If we allow resource dynamics to be much faster than those of the competing species, we can apply an adiabatic approximation and set the latter equations to zero:
\begin{equation}
    m_\alpha(K_\alpha-R_\alpha)-\sum_j c_{\alpha j}^T N_j=0 \ ,
    \label{stationary_R}
\end{equation}
where $m_\alpha$ denotes the ratio between the growth rate and the carrying capacity for each $\alpha$.
Equation (\ref{stationary_R}) implies a closed-form expression for the resources as a function of species' abundances
\begin{equation}
    R_\alpha=K_\alpha -\frac{1}{m_\alpha} \sum_j c_{j \alpha} N_j \ .
\end{equation}
As a consequence, the former dynamic turns out to be:
\begin{equation}
    \frac{\dot{N}_i}{N_i}=\sum_\delta \left[ c_{i \delta} \left( K_\delta -\frac{1}{m_\delta} \sum_j c_{j \delta} N_j \right)\right]-\chi_i=\left( \sum_\delta c_{i \delta} K_\delta -\chi_i\right)-\sum_\delta \left(\frac{1}{m_\delta}\right) \sum_j c_{i \delta} c_{\delta j}^{T} N_j \ .
    \label{dynamics_reduced}
\end{equation}
It can be rephrased in a more compact form:
\begin{equation}
 \frac{\dot{N}_i}{N_i}=\mathcal{U}_i-A_{ii}N_i-\sum_{j\neq i} A_{ij}N_j
\end{equation}
by slightly shaping the interaction matrix 
\begin{equation}
     A_{ij} = \sum_{\delta} c_{i,\delta} c^{T}_{\delta,j} \ .
\end{equation}
Note that we have made use of two assumptions: i) we have assumed $m_\delta=m=1$, namely species-independent parameters and constant over time; ii) 
we have split the sum over $j$ in Eq. (\ref{dynamics_reduced}) into a diagonal and off-diagonal contribution, and eventually swapped the summation over resources and species. In this rephrasing, the interactions are written as a scalar product between uptake profiles, which allows us to recover the Generalized Lotka-Volterra equations with symmetric interactions $A_{ij}=A_{ji}$ .
A detailed derivation can be found in the SI of \cite{pasqualini_microbiomes_2024}.

\subsection{Cavity method for the GLV model: zero-noise picture}
\label{cavity}

One possibility for investigating how a system reacts to external perturbations is to apply linear stability analysis, linearizing the dynamical equations around their fixed point. This approach reflects May's original intuition. 

However, to account for more complex scenarios, especially those featuring multiple equilibria or persistent fluctuations, one can use the \emph{cavity method}, a powerful technique in the analysis of disordered systems. Therefore, we consider the multi-species dynamics for the species abundances $N_i$, without any demographic or environmental noise:
\begin{equation}
    \frac{dN_i}{dt}=r_i N_i\left(K_i-N_i-\sum_{j \neq i} A_{ij} N_j\right) \ .
    \label{dynamics_zero}
\end{equation}
As in previous frameworks, the index $i=1,2,...,S$ runs over the total number of species in the pool. Positive interaction coefficients $A_{ij}>0$ denote competition, whereas negative coefficients stand for mutualism. For simplicity, we shall set the growth rates $r_i=r=1$ and the carrying capacities $K_i=K$ (upon rescaling the other parameters). The analysis can nevertheless be extended in the case of random carrying capacities \cite{biroli_marginally_2018}, providing only a quantitative change in terms of the resulting phase diagram. 

Our goal is to characterize the steady-state solution, defined by $\frac{dN_i}{dt}=0$, where the coefficients $A_{ij}$ are extracted from a suitably defined random matrix. Let us consider then:
\begin{equation}
N_i \left(K-N_i-\sum_{j \neq i} A_{ij} N_j\right)=0     
\end{equation}
where, similarly to the MacArthur model, $\langle A_{ij}\rangle =\mu/S$ and $\langle A_{ij}^2\rangle_c=\sigma^2/S$. Such scaling is essential to ensure a well-defined thermodynamic limit, that is the quantity $\sum_j A_{ij}N_j^{*}$ must be independent of the system size. Therefore, we assume that the elements scale as
\begin{equation}
A_{ij}=\frac{\mu}{S}+\sigma a_{ij}
    \label{expression_A}
\end{equation}
where the first and the second moments of the Gaussian variable $a_{ij}$ read
\begin{equation}
    \langle a_{ij}\rangle=0 \ , \hspace{0.3cm} \langle a_{ij} a_{kl}\rangle=\frac{1}{S} \delta_{ik}\delta_{jl}+\frac{\rho}{S} \delta_{il} \delta_{jk} \ .
\end{equation}
However, writing down a closed-form set of equations does not require the interactions to be Gaussian. The only condition is that the first two moments are finite, and the distribution does not exhibit long tails.

In the cavity formalism, a new species is added to the ecosystem with mutual interactions $A_{0j}$ and $A_{j0}$ with the rest of the community. By construction, the different trajectories $\lbrace N_i(t)\rbrace$ evolving according to the dynamics  (\ref{dynamics_zero}) are independent of such interaction coefficients, which allow us to use central-limit arguments.
The dynamics of the former $S$ species can be written as
\begin{equation}
N_i\left( K-N_i-\sum_j A_{ij}N_j -A_{i0}N_0 \right)=0 \ ,
\end{equation}
whereas for the new one, we have:
\begin{equation}
N_0 \left (K-N_0-\sum_j A_{0j}N_j \right)=0 \ .
\label{dynamics_zero}
\end{equation}
For compactness, we have dropped the asterisk in the steady-state solution. 
As we have seen for the consumer-resource model, adding a new species is equivalent to slightly perturbing the carrying capacity by an amount $\delta K_i$. 
Let us consider the local susceptibilities $\chi_{ij}=\frac{\partial N_i}{\partial K_j}$, and apply the linear response theory to rewrite the stationary solution in perturbation theory:
\begin{equation}
N_i=N_{i \backslash 0} -\sum_{j} \chi_{ij} A_{j0}N_0 \ .
\end{equation}
The term $N_{i \backslash 0}$ represents the abundance in the absence of the $0$-th species, which gives rise to the \emph{cavity} designation.
Since $A_{0i} \sim O\left(\frac{1}{S}\right)$, the term $A_{i0}N_0$ can be treated as a small perturbation.
Plugging the expression above into Eq. (\ref{dynamics_zero}) for the $0$-th species, we end up with
\begin{equation}
    N_0 \left[K-N_0 -\sum_j A_{0j}\left( N_{j \backslash0} -\sum_i \chi_{ji} A_{i0}N_0 \right) \right]=0 \ .
\end{equation}
Solving the equation for $N_0$, and excluding the trivial vanishing solution, we obtain to the leading order:
\begin{equation}
    N_0=\frac{K-\sum_j \sigma a_{0j}N_{j \backslash 0}-\mu \langle N \rangle}{1-\sigma^2 \sum_{ij}\chi_{ij}a_{0j}a_{i0}} \ .
\end{equation}

According to the central limit theorem, one can claim that the contribution $\sum_i a_{0i} \chi_{ii} a_{i0}$ will converge to its expectation value.
%By replacing $a_{0j} a_{i0}$ with their expectation values, 
We can therefore rewrite the term at the denominator as $\rho \frac{1}{S} \sum_{j} \chi_{jj}=\rho\chi$ yielding
\begin{equation}
N_0=\text{max}\left[0, \frac{K-\sum_j \sigma a_{0j}N_{j \backslash 0} -\mu \langle N\rangle}{1-\sigma^2 \rho \chi} \right]
\end{equation}
which is required to be non-negative, otherwise the solution does not exist.
 Then, one need to evaluate the scaling of the typical fluctuations: the non-diagonal term is associated to corrections $O(S^{-1/2})$, which become subleading in the thermodynamic limit.

As a next fundamental hypothesis to close the equations on $N_0$ we need to ask for \textbf{self-averaging} property in the thermodynamic limit, that implies:
\begin{equation}
\begin{split}
& \biggl \langle \sum_{j=1}^{S} \sigma a_{0 j}  N_{j \backslash 0} \biggr \rangle= S \sigma \langle a_{0j}  \rangle \langle N \rangle =0 \ ,\\
& \Biggl \langle \left( \sum_{j=1}^{S} \sigma a_{0 j} N_{j \backslash 0} \right)^2 \Biggr \rangle= \sigma^2 \sum_{j,k} \langle a_{0j} a_{0k} \rangle N_{j \backslash 0} N_{k \backslash 0}=\sigma^2 \langle N^2 \rangle 
\end{split}
\end{equation}
where we have introduced $\frac{1}{S} \sum_{j} N_{j \backslash 0} =\langle N_0 \rangle=\langle N \rangle$, $\frac{1}{S} \sum_{j} N_{j \backslash 0}^2 =\langle N_0^2\rangle= \langle N^2 \rangle$.
By gathering all ingredients together and introducing a standard normal variable $z$, we obtain the distribution of the solutions for $N_0$:
\begin{equation}
N_0=\text{max} \left[0, \frac{K-\mu \langle N\rangle+\sqrt{\sigma^2 \langle N^2 \rangle} z}{1-\sigma^2 \rho \chi} \right]
\end{equation}
which reflects the typical shape of a Gaussian distribution with mean $(K-\mu \langle N \rangle)/(1-\sigma^2 \rho \chi)$ and variance $\sigma^2\langle N^2 \rangle/(1-\sigma^2 \rho\chi)^2$.
To the leading order in $1/S$, the resulting distribution of the typical species abundance is a truncated (positive-definite) Gaussian law. It inherently implies the resolution of the following set of self-consistent equations:
\begin{equation}
\langle N\rangle =\frac{1}{S} \sum_i N_{i \backslash 0} \ , \hspace{0.6cm} \langle N^2 \rangle=\frac{1}{S} \sum_i  N_{i \backslash 0}^2 \ ,
\end{equation}
\begin{equation}
 \chi=\biggl \langle \frac{\partial N_0}{\partial \xi_0}\biggr \rangle =\frac{\phi}{1-\sigma^2 \rho \chi}  \ , \hspace{0.6cm} \phi=\frac{1}{S} \sum_i \theta(N_0) \ .
\end{equation}
where the susceptibility essentially accounts for the single-species response function to a slight perturbation, for instance in the carrying capacity or the growth rate. It can be recast in terms of the fraction of surviving species $\phi$ and $\chi$ itself. The symbol $\theta(\cdot)$ denotes the Heaviside step function. See also \cite{cui_houches_2024} for an alternative derivation in a few benchmark models.

\section{High-dimensional random GLV equations with stochasticity}

We consider the following dynamical equations for the evolution of the species abundance $N_i$ at time $t$, where the index $i$ runs over the total number of species $i=1,...,S$:
\begin{equation}
\begin{split}
\frac{d N_i}{dt}= & N_i \left[(K_i-N_i) -\sum_{j,(j\neq i)} A_{ij}N_j \right] +\eta_i(t)+\lambda_i =\\
   =& N_i \left[-\nabla_{N_i} V_i(N_i) -\sum_{j, (j \neq i)} A_{ij} N_j \right] +\eta_i(t) +\lambda_i \ ,
    \label{dynamical_eq}
    \end{split}
\end{equation}
$N_i(t)$ is interpreted as a relative species abundance of species $i$ at time $t$, meaning that the population is actually normalized by the total number of individuals populating the ecosystem in the absence of interactions (deterministic scenario, i.e. $A_{ij}=0$). 
%The parameter $\rho_i \equiv r_i/K_i$ denotes the ratio between the single-species growth rate and the carrying capacity. 

To account for demographic fluctuations, we introduce the variable $\eta_i(t)$, a white Gaussian noise with zero mean and variance $\langle \eta_i(t)\eta_j(t')\rangle=2 T N_i(t)\delta_{ij} \delta(t-t')$,  where $T$ represents the noise amplitude. 
The amplitude $T$ turns out to be inversely proportional to the total number of individuals in the pool: hence, the larger the system size, the smaller the demographic noise strength. 
We specifically follow Itō's convention for the stochastic part: the multiplicative nature of the equations ensures that for zero immigration ($\lambda_i=0$) -- if the absorbing state $N_i=0$
is reached -- the population goes extinct and remains there at all later times. 
Alternatively, to properly study the interplay with an intrinsic source of fluctuations, one can introduce a species-independent immigration parameter $\lambda$, therefore
 preventing overall extinctions in a finite time.

\subsection{Itō's versus Stratonovich prescription}
Equation (\ref{dynamical_eq}) is meaningless without selecting a proper discretization for the noise.
If one adopts Itō's prescription, the time derivative of a generic observable $\mathcal{O}$ can be expressed as
\begin{equation}
\frac{d}{dt} \langle \mathcal{O} ( \lbrace N_j \rbrace) \rangle = \langle \sum_i \frac{\partial \mathcal{O}}{\partial N_i} \frac{ d N_i}{d t} \rangle + T \langle \sum_{i} \frac{\partial^2 \mathcal{O}}{\partial N_i^2} N_i \rangle \ ,
\end{equation}
where  $\langle \cdot \rangle$ stands for the average over the probability distribution $P(\lbrace N_j \rbrace,t)$, i.e. over the thermal noise. 

Then, in the symmetric interaction case (for $A_{ij}=A_{ji}$), one can show that the multi-species dynamics admit an invariant equilibrium-like probability distribution \cite{biroli_marginally_2018, altieri_properties_2021}. 
By plugging the dynamical equation (\ref{dynamical_eq}) in the above expression, we obtain: 
{\small{
\begin{equation}
\frac{d}{dt} \langle \mathcal{O}(\lbrace N_j \rbrace) \rangle = 
 \biggl \langle \sum_i \frac{\partial \mathcal{O}}{\partial N_i} \left[ - N_i \nabla_{N_i} V_i(N_i) - N_i \sum_j A_{ij} N_j +\lambda \right] \biggr \rangle + T \langle \sum_i \frac{\partial^2 \mathcal{O}}{\partial N_i^2} N_i \rangle \ .
\end{equation}
}}
The three terms in parenthesis can be collected together and denoted as $\tilde{F}(\lbrace N_j \rbrace)$ allowing us to rewrite the time evolution of the average operator as
\begin{equation}
\frac{d}{d t} \langle \mathcal{O} (\lbrace N_j \rbrace) \rangle = \int \prod_i d N_i P (\lbrace N_j \rbrace, t) \left[ \sum_i \frac{\partial \mathcal{O}}{\partial N_i} \tilde{F}(\lbrace N_j \rbrace) + T \sum_i \frac{\partial ^2\mathcal{O}}{\partial N_i^2} N_i \right] \ ,
\end{equation}
from which the resulting equation for $\langle \mathcal{O}(\lbrace N_j \rbrace) \rangle$ can easily be obtained by integration by parts.
%\begin{equation}
%\frac{d}{d t} \langle \mathcal{A} (\lbrace N_j \rbrace) \rangle =\int \prod_i d N_i \mathcal{A} (\lbrace N_j \rbrace) \sum_i \left[ -\frac{\partial}{\partial N_i} \left[\tilde{F}(\lbrace N_j \rbrace) P(\lbrace N_j \rbrace,t) \right] + T \frac{\partial^2}{\partial N_i^2} \left[ P (\lbrace N_j \rbrace, t) N_i \right] \right] \ .\end{equation}
In the same spirit, the dynamical equation for the probability distribution reads:
\begin{equation}
\frac{\partial P ( \lbrace N_j \rbrace, t) }{\partial t} =\sum_i \left[ -\frac{\partial}{\partial N_i} \left[\tilde{F}(\lbrace N_j \rbrace) P(\lbrace N_j \rbrace,t) \right] + T \frac{\partial^2}{\partial N_i^2} \left[ P (\lbrace N_j \rbrace, t) N_i \right] \right]
\end{equation}
from which, by imposing the R.H.S. to be zero, we obtain the stationary probability distribution. Therefore, we ask the invariant probability distribution to scale as $P \propto \exp(-\beta H)$, with $H$ an effective energy function, and inverse temperature $\beta=1/T$.
This implies $
\frac{1}{P}\frac{\partial P}{\partial N_i}=-\frac{1}{T}\frac{\partial H} {\partial N_i}$ from which, by integrating over $N_i$, we can deduce the associated energy function valid in the symmetric interaction case:
\begin{equation}
    H=\sum_{i} V_i(N_i) +\sum_{i<j}A_{ij}N_i N_j +(T-\lambda) \sum_i \ln N_i \ .
    \label{H_LV}
\end{equation}
A complete derivation can be found for other instances of conservative dynamics, for instance in the case of a non-logistic self-regulation contribution (see \cite{Altieri-Biroli2022}). 

In other words, the original dynamical process in Eq. (\ref{dynamical_eq}) describes the time evolution of an interacting ecosystem whose thermodynamics is defined by Eq.  (\ref{H_LV}).
Note that -- at variance with the simplest scenario corresponding to the analysis in the limit $T \rightarrow 0$ and $\lambda \rightarrow 0^{+}$ -- introducing a small but finite immigration results in a qualitative behavioral change. The immigration parameter ensures that no species will go extinct thus replacing the concept of indefinite extinction with the concept \emph{everything is everywhere} \cite{grilli_macroecological_2020}.

Conversely, Stratonovich discretization has the advantage of preserving invariance under time reversal. However, due to the multiplicative nature of the noise, the two discretization schemes are not formally equivalent and necessitate the introduction of an additional drift term:
\begin{equation}
    \eta \sqrt{N} \rightarrow \eta \sqrt{N} -\frac{1}{2}\frac{\sqrt{2T}}{2 \sqrt{N}} \sqrt{2TN} = \eta \sqrt{N}-\frac{T}{2} \ .
\end{equation}

\subsection{Disordered free energy: replica method}

The disorder average of the free energy can be computed through the replica method \cite{MPV}, a powerful technique originally introduced for dealing with spin-glass problems. It relies on the identity
\begin{equation}
	-\beta F=\overline{\ln Z}=\lim_{n\to 0}\frac{\ln \overline{Z^n}}{n} \ .
\end{equation}
where one first computes the replicated partition function for integer values of $n$, and only at the final stage considers the analytical continuation $n\to 0$.
Despite being thermodynamically independent, replicas turn out to be correlated because they are taken over the same realization of the disorder. Therefore, integrating over the disorder introduces an effective coupling between them. 
%The similarity between their states is measured by the overlap

%Studying the overlap between them allows us to understand the structure of the phase space

Let us thus proceed with the computation of $\overline{Z^n}$:
\begin{equation}
	\overline{Z^n}=\int\prod_{i<j}d A_{ij}\exp\left(-\sum_{i<j}\frac{(A_{ij}-\mu/S)^2}{2\sigma^2/S}\right)\int\prod_{a=1}^n\prod_idN_i^a\exp\left\{-\beta \sum_a H(\{N_i^a\})\right\}
\end{equation}
By performing a Gaussian integration over the random variables $A_{ij}$ with finite mean and variance, we obtain:
\begin{equation}
		\overline{Z^n}=\int\prod_{a=1}^n\prod_idN_i^a\exp\Bigg\{ \sum_{i<j}\frac{\sigma^2}{2S}\left(\beta \sum_aN_i^aN_j^a\right)^2 -\beta\sum_a\Bigg[\sum_i \left[V_i(N_i)+(T-\lambda)\ln(N_i) \right] + \frac{\mu}{S}\sum_{i<j}N_i^aN_j^a \Bigg] \Bigg\}
\end{equation}
To decouple the first term, we thus introduce the following quantities corresponding respectively to the \emph{overlap matrix}  between two replicas $a$, $b$ of the reference system and the \emph{mean abundance}:
\begin{equation}
Q_{ab}=\frac{1}{S}\sum_{i=1}^{S}N_i^a N_i^b \ , \hspace{0.5cm} H_a=\frac{1}{S}\sum_{i=1}^S N_i^a
\end{equation}
The replicated partition function can eventually be rewritten in terms of the new order parameters:
\begin{equation}
	\overline{Z^n}=\int\prod_{a\leq b}dQ_{ab}\prod_adH_a\exp\left\{S\mathcal{A}\left(\{Q_{ab}, H_a\}\right)\right\}
\end{equation}
where the free-energy action $\mathcal{A}$ reads:
\begin{equation}
		\mathcal{A}\left(\{Q_{ab}, Q_{aa}, H_a\}\right)
		=-\frac{1}{2}\sigma^2\beta^2\left(\sum_{a<b}Q_{ab}^2+\frac{1}{2}\sum_aQ_{aa}^2\right)+\frac{\beta\mu}{2}\sum_aH_a^2+\frac{1}{S}\sum_i\ln Z_i(\{Q_{ab}, H_a\}) \ .
        \label{action_overlap}
\end{equation}
Alternatively, one can recognize a Hubbard-Stratonovich transformation structure, which in the case of a multivariate Gaussian integral reads:
%in the scalar case yields
%{\color{magenta}\begin{equation}
%    \exp \left( \frac{x^ 2}{2 \sigma^2} \right)=\frac{1}{\sqrt{2 \pi \sigma^2}} \int_{- \infty}^{\infty} \exp \left[ -\frac{y^2}{2\sigma^2}+\frac{xy}{\sigma^2} \right] dy
%\end{equation}
%}
\begin{equation}
    \sqrt{\frac{(2 \pi)^N}{\det \hat{A}}}\exp \left( -\frac{\textbf{b}^{T}\hat{A}\textbf{b}}{2} \right)= \int_{-\infty}^{\infty} \exp \left[ -\frac{\textbf{y}^{T}\hat{A}^{-1} \textbf{y}}{2}+\textbf{b}^{T}\cdot\textbf{y} \right] d\textbf{y} \ ,
\end{equation}
which we apply to $\boldsymbol{y}=Q_{ab}$ (and $H_a$, in turn).

The last piece in Eq. (\ref{action_overlap}), i.e. the effective partition function for the one-species abundances in different replicas, reads:
\begin{equation}
	Z_i=\int\prod_adN_i^a\exp\left[-\beta H_{\text{eff}}(\{N_i^a\},\{Q_{ab}, H_a\})\right]
\end{equation}
whose associated effective Hamiltonian is
\begin{equation}
		H_{\text{eff}}=-\beta\sigma^2\left(\sum_{a<b}N_i^aN_i^bQ_{ab}+\frac{1}{2}\sum_a(N_i^a)^2Q_{aa}\right)+\sum_a\bigg(\mu H_aN_i^a+V(N_i^a)+(T-\lambda)\ln N_i^a\bigg) \ .
\end{equation}
\vspace{0.3cm}

\subsection{Saddle-point evaluation of the free energy}

In the $S\to\infty$ limit only the values of $Q_{ab}$ and $H_a$ that extremize the action will contribute\footnote{$H_a$ would actually need to be integrated over the imaginary axis; deforming the integration contour and using the method of steepest descent would then yield the stated result.}:
\begin{equation}
	\frac{\ln \overline{Z^n}}{S}=\mathcal{A}(\{Q_{ab}^*, H_a^*\}) \ .
\end{equation}
Therefore, based on the stationarity condition, one obtains:
\begin{align}
	\label{eq:selfconsab}
\begin{aligned}
	\frac{\partial\mathcal{A}}{\partial Q_{ab}}=0\\
	\frac{\partial\mathcal{A}}{\partial H_{a}}=0
\end{aligned}
	\qquad\longrightarrow\qquad
	\begin{aligned}
		Q^*_{ab}=
		\frac{1}{S}\sum_i\langle N_i^{a} N_i^{b}\rangle\\
		H^*_{a}=
		\frac{1}{S}\sum_i\langle N_i^{a} \rangle
	\end{aligned}
\end{align}
where the brackets indicate the thermal average over the resulting Hamiltonian $H_\text{eff}$. It depends on $Q_{ab}$ and $H_a$ requiring the explicit integration or, in a more straightforward way, the resolution of a set of self-consistency equations.
As it is for spin-glass models, $Q_{ab}$ represents the overlap matrix between two replicated configurations, in the presence of an external field, denoted by $H_a$.

\subsection{The replica symmetric solution}

\begin{figure}[h]
\centering
\includegraphics[scale=0.45]{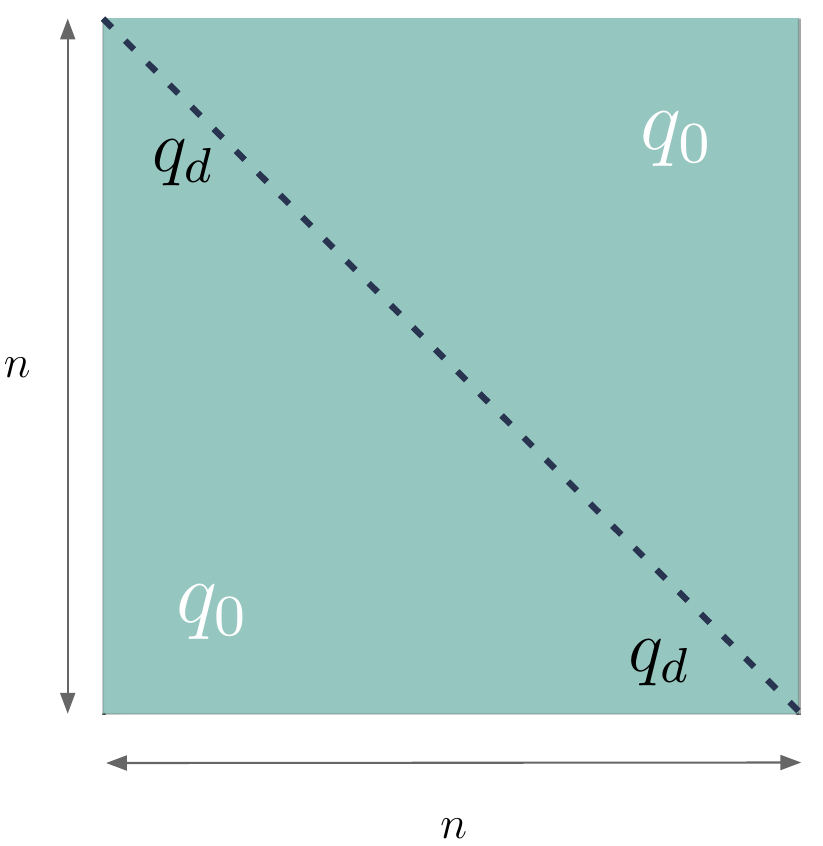}
\caption{Replica-Symmetric (RS) ansatz for the overlap matrix $Q_{ab}$, which is parametrized by two values only, the diagonal term (self-overlap) and the off-diagonal contribution (inter-state overlap).}
\label{matrixRS}
\end{figure}

Solving these high-dimensional equations without guessing a specific form for the overlap matrix $Q_{ab}$ and the mean abundance $H_a$ seems impractical. 
To address this challenging task, one can initiate the analysis by assuming a replica-symmetric (RS) ansatz.
Since the action is symmetric under the exchange of replica indices, we can first claim that the solution respects this symmetry. Therefore, all fields must be equal and the overlap can only take two values: $q_0$ for the overlap between different replicas, and $q_d>q_0$ for the self-overlap (see Fig. \ref{matrixRS}).
\begin{align}
	\begin{aligned}
		Q_{ab}=q_0\\
		Q_{aa}=q_d\\
		H_a=h
	\end{aligned}
	&&
	\begin{aligned}
		a\neq b\\
		a=b\\
		\forall a
	\end{aligned}
    \label{overlap-RS}
\end{align}

The stability of the solution will be eventually checked by computing the Hessian of the free energy, and notably its leading eigenvalue, the so-called replicon eigenvalue \cite{Giardina-DeDom, MPV, altieri-baity2024}.
If there were multiple equilibria, two distinct replicas would have a different overlap according to whether they were in the same state or not.
Since here distinct replicas always have overlap $q_0$, this ansatz corresponds to assuming that there is a unique equilibrium state.
The "size" of this state is characterized by the overlap between different replicas $q_0$: if it is large ($q_0\lesssim q_d$) the configurations in the same state are very similar, so that the state is very localized in phase space, if it is small ($q_0\ll q_d$) the state is very wide. 
We can insert it in the action and in the effective Hamiltonian:
\begin{equation}
	\mathcal{A}(q_d,q_0,h)=-\frac{1}{2}\sigma^2\beta^2\left(\frac{n(n-1)}{2}q_0^2+\frac{n}{2}q_d^2\right)+\frac{\beta\mu n}{2}h^2+\frac{1}{S}\sum_i\ln Z_{i}(q_d, q_0, h)
\end{equation}
\begin{equation}
		H_{eff}=-\frac{\beta\sigma^2}{2}\left(q_0\left(\sum_{a}N_i^a\right)^2+(q_d-q_0)\sum_a(N_i^a)^2\right)+\sum_a\left(\mu hN_i^a+V(N_i^a)+(T-\lambda)\ln N_i^a\right)
\end{equation}
To decouple the different replicas we introduce a Gaussian integration over an auxiliary variable $z_i$, thus obtaining:
\begin{equation}
	Z_i=\int_{-\infty}^{+\infty}\frac{dz_i}{\sqrt{2\pi}}e^{-z_i^2}\left[\int dN_i\exp\left\{-\beta H^{RS}_\text{eff}(N_i; q_d, q_0, h;z)\right\}\right]^n
\end{equation}
\begin{equation}
	H^{RS}_\text{eff}=-\beta\sigma^2\frac{q_d-q_0}{2}N_i^2+(\mu h-z\sqrt{q_0}
	\sigma)N_i+V(N_i)+(T-\lambda)\ln N_i
\end{equation}

One of these terms is proportional to a fluctuating field $z$, which accounts for the randomness of the interactions.
Considering the stationary condition, performing the analytic continuation for $n \to 0$ and taking into account that after the integration over the fluctuating field $z$ all species are equivalent, the self-consistent equations (\ref{eq:selfconsab}) become:
\begin{align}
	\begin{aligned}
	\label{eq:selfconsa}h &=
		\int \mathcal{D}z \left(\frac{\int_0^\infty dN e^{-\beta H^{RS}_\text{eff}(q_0, q_d, h, z)}N}{\int_0^\infty dN e^{-\beta H^{RS}_\text{eff}(q_0, q_d, h, z)}}\right)= \overline{\langle N \rangle} \\
	q_d & =\int \mathcal{D}z \left(\frac{\int_0^\infty dN e^{-\beta H^{RS}_\text{eff}(q_0, q_d, h, z)}N^2}{\int_0^\infty dN e^{-\beta H^{RS}_\text{eff}(q_0, q_d, h, z)}}\right) = \overline{\langle N^2 \rangle}\\
	q_0 &=\int \mathcal{D}z \left(\frac{\int_0^\infty dN e^{-\beta H^{RS}_\text{eff}(q_0, q_d, h, z)}N}{\int_0^\infty dN e^{-\beta H^{RS}_\text{eff}(q_0, q_d, h, z)}}\right)^2= \overline{\langle N \rangle^2} 
	\end{aligned}
\end{align}
where we used the calligraphic notation for the Gaussian integral in $z$: $\int\mathcal{D}z\equiv\int_{-\infty}^{\infty}\frac{dz}{\sqrt{2\pi}}e^{-z^2/2}$.
The brackets indicate the average over the Boltzmann distribution under the effective Hamiltonian, while the overbar denotes the average over the disorder, represented by the Gaussian unit-variance variable $z$.
The two averages coincide with a single species abundance's thermal and disorder average, respectively.
Given the resulting values of the above set of equations, one can eventually compute the disordered free energy as:
\begin{equation}
	f=\frac{F}{S}=-\lim_{n\to0}\frac{\ln \overline{Z^n}}{n \beta S}= - \lim_{n\to 0}\frac{\mathcal{A}(h, q_0, q_d)}{\beta n} \ .
\end{equation}

Solving these equations analytically at finite $\beta$ is impractical. It is thus necessary to employ a numerical scheme that, given an initial guess, allows us to solve them iteratively. 
\vspace{0.2cm}

\begin{mybox}
\par\noindent\rule{\textwidth}{0.8pt}

\texttt{Algorithmic protocol}
\par\noindent\rule{\textwidth}{0.8pt}

\begin{itemize}
\item{Initialize the mean abundance $h$, and the two overlaps $q_d$, $q_0$ at the initial time;}
\item{Solve Eqs. (\ref{eq:selfconsa}) iteratively upon progressively increasing $\beta=1/T$. The introduction of a damping parameter -- for instance $\alpha=0.1$ -- can be beneficial to facilitate the convergence.

Therefore, by looking at the first equation, one should consider: \\
$h^{t} \leftarrow \alpha\int \mathcal{D} z \frac{ \int_{N_c}^{\infty} d N e^{-\beta H_{\text{RS}}(q_0^{t-1},q_d^{t-1},h^{t-1},z)} N}{ \int_{N_c}^{\infty} d N e^{-\beta H_{\text{RS}}(q_0^{t-1},q_d^{t-1},h^{t-1},z)}}+(1-\alpha)h^{t-1}$ .} 
\item{The algorithm is expected to converge if the error between the $(t-1)$-value and $t$-value $ \le \epsilon$, a given arbitrary precision $\epsilon$, for each of the three order parameters.}
\end{itemize}

\end{mybox}

\vspace{0.1cm}

\subsection{Numerical interpretation of the order parameters}

To validate the replica symmetric ansatz, one can simulate the dynamical system defined in Eq. (\ref{dynamical_eq}).
To generate a single ecosystem realization:
\begin{itemize}
    \item First, one needs to sample the $S \times S$ interaction matrix with rescaled parameters $(\tilde{\mu}, \tilde{\sigma}^2)$;
    \item One sets up the initial conditions $N_i(t=0)$, by assuming for instance a uniform distribution in $[0,1]$;
    \item Then, demographic fluctuations are sampled from a white-noise distribution with amplitude $T$.
\end{itemize}
The dynamical equations can then be integrated deterministically up to a given $t_\text{max}$, which denotes the maximal extent of the simulation. Also, to analyze variability, it is beneficial to perform the simulation for $N_\text{samples}$ independent realizations, each obtained by varying the sources of randomness. This
process produces a dataset $\lbrace N_i^p(t) \rbrace_{t=0,...,t_\text{max};i=1,...,S; p=1,...,N_\text{sample}}$.
Eventually averaging over all these different contributions, one can have access to the dynamical correlator:
\begin{equation}
\mathbb{E}[N(t)N(t')]= \frac{1}{S} \frac{1}{N_\text{sample}} \sum_{i=1}^S \sum_{p=1}^{N_\text{sample}}N_i^{p}(t) N_i^p(t') \ ,
\end{equation}
which, for sufficiently large samples ($N \simeq 50-60$), shows a convergence in law. Moreover, at sufficiently large times, the correlator appears to satisfy the time translational invariance (TTI) property being a function only of the time difference: $\mathbb{E}[N(t)N(t')]= C(t-t')$, $\forall t \ge t' > t_\text{wait}$. The waiting time depends on the parameters' choice, notably $(\sigma, T)$; however, in the single equilibrium phase, we are guaranteed that the dynamics converge to the TTI state and $t_\text{wait}$ turns out to be typically $10^2$. 
\begin{figure}[htbp]
  \begin{minipage}[t]{0.46\textwidth}   \includegraphics[align=t,width=\textwidth]{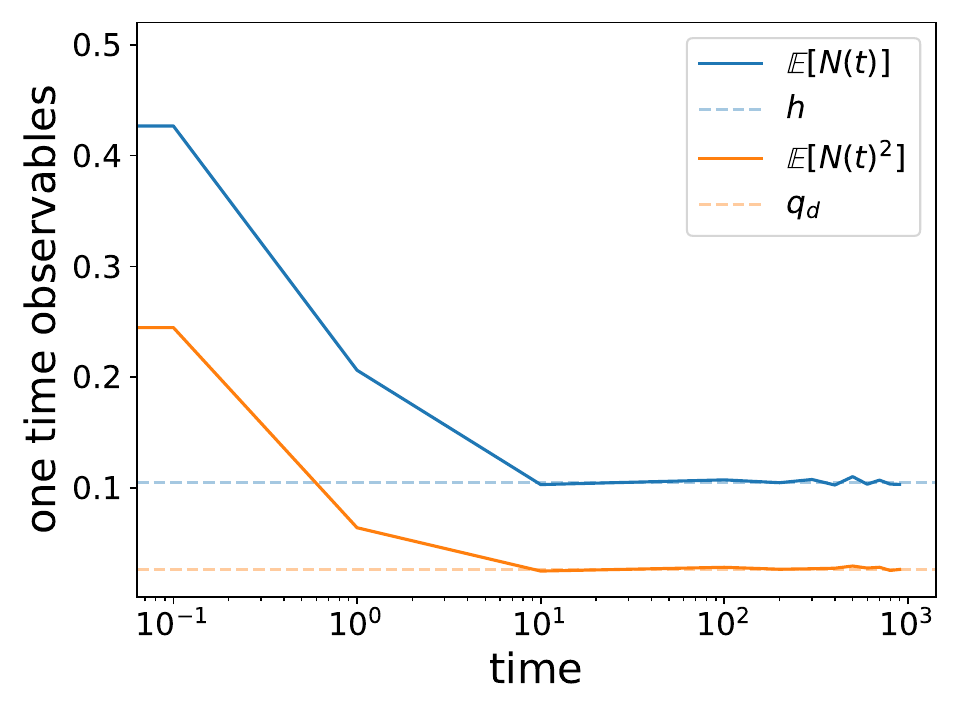}  %\caption{Self-overlap and mean abundance in the RS regime. The dashed line corresponds to the theoretical prediction; the full line results from the integration of Dynamical Mean-Field Theory equations.}
  \end{minipage}
   \begin{minipage}[t]{0.54\textwidth}
The (left) Figure shows the numerical comparison for two characteristic one-time observables: the self-overlap $q_d$ and the mean abundance $h$ in the single-equilibrium (RS) regime. The dashed line corresponds to the theoretical prediction; the full line results from the integration of Dynamical Mean-Field Theory equations. For more details see \cite{roy_numerical_2019, altieri_properties_2021}.
  \end{minipage}
\end{figure}

We eventually find:
\begin{equation}
h=\mathbb{E}[N(t)] \ , \hspace{0.3cm} q_d=C(0)=\mathbb[N(t)^2] \ , \hspace{0.3cm} q_0=\lim_{\tau \rightarrow \infty} C(\tau) \simeq \mathbb{E}[N(t)N(t_\text{max})] \ ,
\end{equation}
where the L.H.S. is computed through the replica method and the R.H.S. provides the numerical comparison. 

\subsection{Stability analysis: Single equilibrium versus multiple equilibria}

In the symmetric interaction case, for $A_{ij}=A_{ji}$, the RS solution becomes unstable at $\sigma_c=\frac{1}{\sqrt{2}}$. In the asymmetric case, the instability bound is slightly modified to take the correlation between asymmetric coefficients into account \cite{Bunin2016}. Without correlation, the transition from the unique equilibrium to multiple attractors turns out to be independent of $\mu$ and lies
on the line $\sigma=\sqrt{2}$. For a generic asymmetry $\rho$, on the other hand, $\sigma_c=\sqrt{2}/(1+\rho)$. Increasing the symmetry shifts the two transitions
towards lower variance and stronger interactions. In this light,
predation-prey relationships may contribute to stabilizing the community.

To investigate the stability of the different phases, we introduce the Hessian matrix of the free energy, which allows us to study the harmonic fluctuations in terms of the matrix $\delta Q_{ab}$. Thanks to the symmetry group properties of the replica space, the diagonalization of the Hessian matrix can be expressed as a function of three different contributions. Following \cite{bray1979, Giardina-DeDom}, we define three sectors: the \emph{longitudinal}, $\lambda_\text{L}$, the \emph{anomalous}, $\lambda_\text{A}$, and the \emph{replicon}, $\lambda_\text{R}$.

An instability leading to the breaking of replica symmetry is signaled by the Hessian matrix, projected on the so-called replicon sector, no longer being strictly positive definite.
If its smallest eigenvalue reaches zero, this signals the onset of an instability, analogous to the de Almeida-Thouless line, as originally identified in the Sherrington-Kirkpatrick model for spin glasses \cite{dealmeida1978}.

Conversely, a zero longitudinal mode provides insights into spinodal points describing how a state opens up along an unstable direction. This phenomenon is particularly relevant for understanding the emergence of \emph{tipping points} and catastrophic collapses, which typically occur through first-order phase transitions.
To be more detailed, we consider the matrix of the second derivatives of the action, as in Eq. (\ref{action_overlap}), differentiated with respect to the overlap matrix:
\begin{equation}
   \mathcal{M}_{abcd} \equiv  -\frac{\partial^2 \mathcal{A}}{\partial Q_{ab}\partial Q_{cd}}=\beta^2 \rho^2 \sigma^2 \left[\delta_{(ab),(cd)}-(\beta^2 \rho^2 \sigma^2) \overline{\langle N^a N^b, N^c N^d \rangle_c}\right] \ .
   \label{Mass}
\end{equation}
The subscript $\langle \cdot \rangle _c$ denotes the connected part of the correlation function. 

It should be noted that the limit $n \rightarrow 0$ implies a negative multiplicity for $Q$s, thus entailing a change in the convexity of the free energy at the stationary point. Therefore, one should evaluate the corrections to the saddle point and select only those solutions with a positive-defined second derivative of the action.

Within the replica formalism, Eq. (\ref{Mass}) can eventually be recast as a function of three different correlators, depending on the choice of the replica indices:
\begin{equation}
   \mathcal{M}_{abcd}= M_{ab,ab} \left( \frac{\delta_{ac}\delta_{bd}+\delta_{ad}\delta_{bc}}{2}\right) +  M_{ab,ac} \left( \frac{\delta_{ac}+\delta_{bd}+\delta_{ad} +\delta_{bc}}{4} \right)+ M_{ab,cd} \ .
\end{equation}
In the specific case of the replicon sector, the associated eigenvalue reads:
\begin{equation}
 \lambda_\text{R}=   (\beta \rho \sigma)^2  \biggl [ 1-(\beta \rho \sigma)^2 \overline{\left( M_{ab,ab}-2M_{ab,ac}+M_{ab,cd} \right) } \biggr ] 
 \end{equation}
where each of these contributions can be rewritten as a function of the species abundance correlators as follows
\begin{equation}
 M_{ab,ab}-2M_{ab,ac}+M_{ab,cd} = \left[ \langle (N^a)^2(N^b)^2\rangle -2 \langle (N^a)^2 N^b N^c \rangle +\langle N^a N^b N^c N^d \rangle \right] \ .
\end{equation}
As usual, the average $\langle \cdot \rangle$ is performed over the effective Hamiltonian, while the disordered one is denoted by $\overline \cdot$. In the presence of a single equilibrium -- namely in the replica-symmetric approximation -- this leads to the following replicon eigenvalue:
\begin{equation}
\lambda_\text{R}=(\beta \rho \sigma)^2 \left[ 1-(\beta \rho \sigma)^2 \overline{ \left(\langle N^2 \rangle - \langle N \rangle^2 \right)^2} \right] \ .
\end{equation}
The averaged difference in the parenthesis accounts for fluctuations between the first and second moment of the species abundances within one state, namely between the diagonal value $q_d$ and the off-diagonal contribution $q_0$ of the overlap matrix. Such a difference can also be interpreted as the response function of the single species to an infinitesimal perturbation.
The identification of a vanishing replicon eigenvalue is intricately linked to the emergence of \emph{marginal states} and diverging susceptibilities.This aspect is also pertinent in the context of out-of-equilibrium aging dynamics.
\begin{figure}
    \centering
    \includegraphics[width=0.7\linewidth]{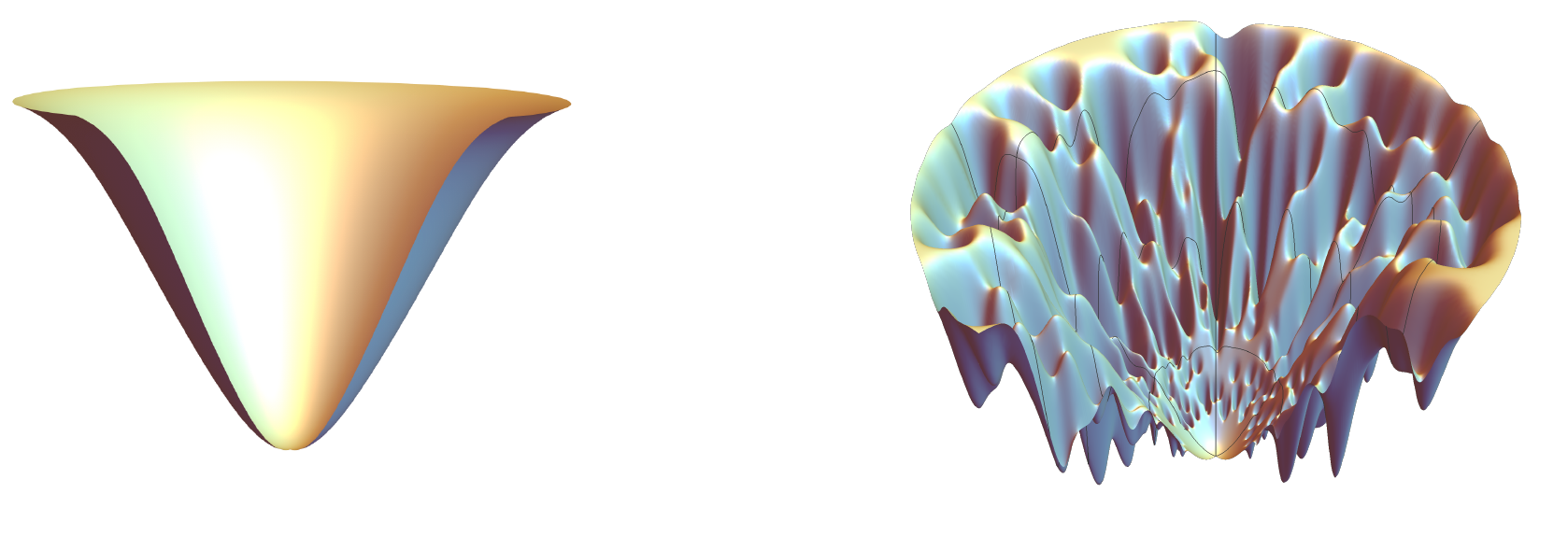}
    \caption{Pictorial representation of a replica-symmetric landscape for the Generalized Lotka-Volterra model, depicted by a single fixed point (left); fractal landscape structure in the low-demographic noise phase, formally captured by a full-RSB solution (right).}
    \label{fig:landscape-single}
\end{figure}
This stability analysis procedure can be repeated as many times as needed, ranging from the 1RSB ansatz to the full-RSB solution. In the 1RSB scenario, a key difference lies in the distinction between the intra-state average, computed within a single replica state, and the inter-state average, computed over a block of replicas of size $m$. In a similar manner, the replicon eigenvalue reads:
\begin{equation}
\lambda_\text{R}^\text{1rsb}=(\beta \rho \sigma)^2 \left[ 1-(\beta \rho \sigma)^2 \overline{ \langle \left(\langle N^2 \rangle_\text{$1$r} - \langle N \rangle^2_\text{$1$r} \right)^2 \rangle_\text{$m$-r}} \right] \ ,
\end{equation}
The point at which the $1$RSB replicon mode vanishes marks a critical phase transition to a more structured phase. 

\section{Derivation of the phase diagram at finite noise}
\label{app_PD}

\begin{figure}[h]
    \centering
\includegraphics[scale=0.55]{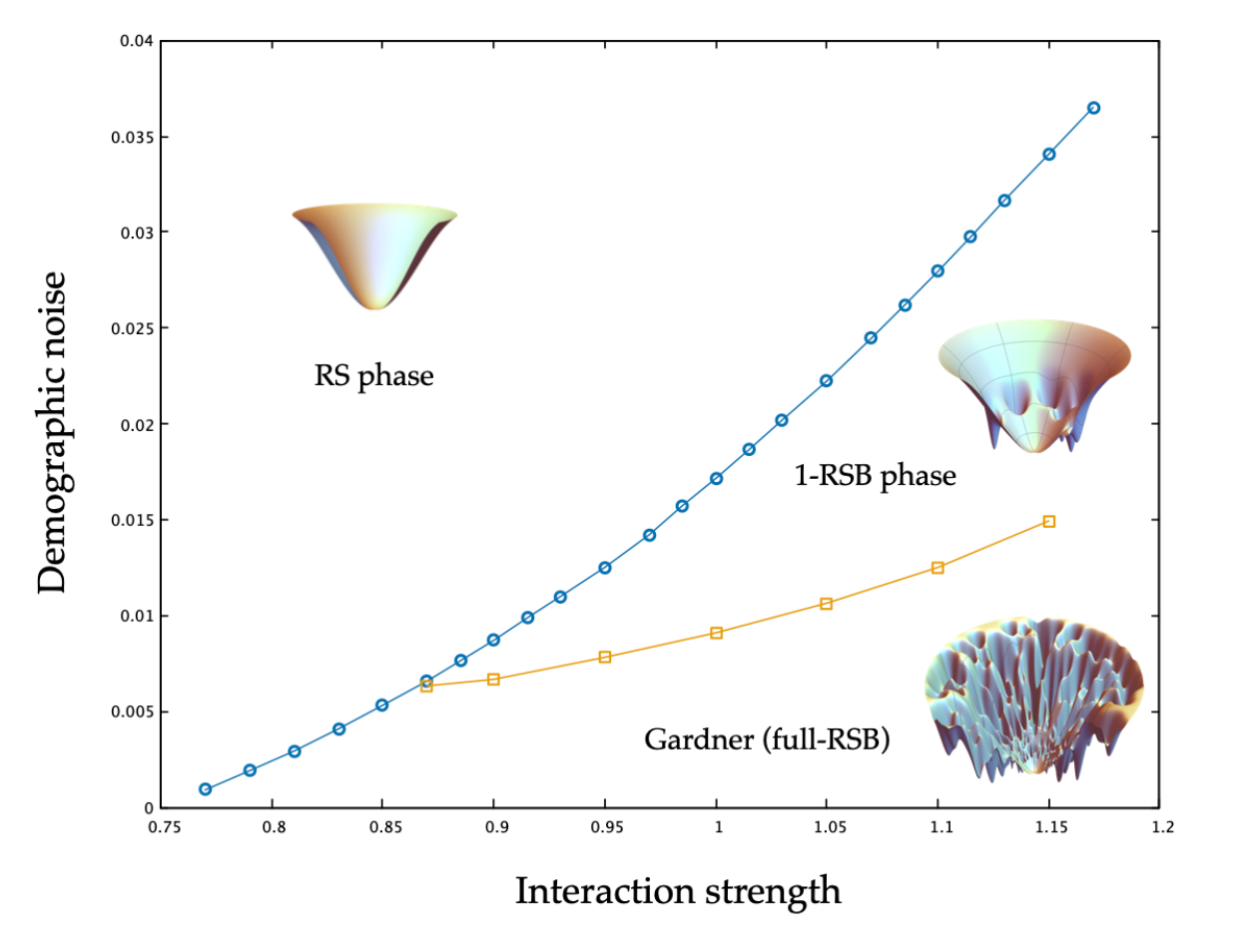}
    \caption{Phase diagram of the GLV model in the presence of finite demographic fluctuations and random symmetric interactions, i.e. $\alpha_{ij}=\alpha_{ji}$.  At sufficiently high demographic noise, the landscape is described by a single fixed point (above the blue line). At low-noise and high-heterogeneity values, the dynamics get trapped in different multiple-equilibria phases, which are captured either by a 1-RSB scenario or a full-RSB one. Figure adapted from \cite{altieri_properties_2021}.}
    \label{fig:PD_log}
\end{figure}

Different regimes associated with increasing complexity can be pointed out by varying the demographic noise amplitude, $T$, and the heterogeneity of the interactions, $\sigma^2$ as pertinent control parameters of the model.
Below the blue line in Fig. (\ref{fig:PD_log}), which reflects the vanishing behavior of RS replicon eigenvalue, the landscape structure is no longer identified by a single equilibrium. It would be depicted, instead, by a more complex, two-level structure, which turns out to be associated with an additional order parameter as well.

Remarkably, upon further decreasing the demographic noise and increasing the variance of the interaction matrix $\sigma^2$, the system undergoes a \emph{Gardner transition} to a marginally stable, amorphous phase, terminology borrowed from glassy physics. In this regime, each state -- also referred to as a basin -- fragments into a fractal hierarchy of sub-basins (\emph{metabasins}), as illustrated in Fig. (\ref{fig:landscape-single}), right panel. As a result, the internal structure of the emerging sub-basins is formally described by the full replica symmetry breaking (full-RSB) solution of the partition function (below the orange line, Fig. (\ref{fig:PD_log})). This regime is characterized by an infinite hierarchy of broken symmetries in the order parameter \cite{MPV}. In other words, the original piecewise function parametrizing the overlap matrix will converge to a continuous function $q(x)$, with $x \in [0,1]$.

Analytical calculations in high-dimensional hard-sphere systems and continuous constraint satisfaction problems have allowed for the prediction of this type of transition, supporting the emergence of a fractal hierarchy \cite{charbonneau2014}. Hence, the infinite-dimensional theory corroborates the hypothesis that an equilibrium glass state undergoes a structural change, transitioning from a normal glass to a marginal one in the low-temperature regime.
Remarkably, for the first time in an ecological context, an analogy between low-temperature glassy systems and complex energy landscapes in the Generalized Lotka-Volterra model has been highlighted in \cite{altieri_properties_2021}, building on prior studies in this direction \cite{biroli_marginally_2018, Bunin2016, altieri_constraint_2019}.
However, what a replica symmetry-breaking scenario precisely yields for biological and ecological communities --  whether linked to a finite or a hierarchical sequence of slower and slower timescales in the dynamics -- represents still an open and intriguing question.

\subsection{Zero-temperature limit of the RS solution}

At $T \rightarrow 0$, the two order parameters tend to become degenerate, i.e. $q_0 \rightarrow q_d$. It is therefore useful to define an auxiliary quantity  $\Delta q= \beta \rho (q_d-q_0)$ that remains finite. In this limit, the integrals are exactly solvable given the vanishing trend of the logarithmic term in the Hamiltonian (\ref{H_LV}) and their consequent Gaussian nature.
\begin{mybox}
The self-consistent equations for the order parameters turn out to be then:
\begin{equation}
\begin{split}
& h=\overline{N^{*}(z)} \ , \\
& q_0=\overline{N^{*}(z)^2}\ , \\
&\Delta q= \frac{\overline{\theta(N^{*}(z)}}{H^{''}_\text{RS}(N^{*}(z))}
\label{sp_zeroT}
\end{split}
\end{equation}
\end{mybox}
where the $\theta(\cdot)$ function in the last expression accounts only for the positive-definite species abundances. The saddle-point approximation eventually yields for the typical value:
\begin{equation}
N^{*}(z)=\text{max}\left[0, \frac{K-\mu h +z \sigma \sqrt{q_0}}{1-\Delta q \sigma^2} \right] \ .
\end{equation}
Equations (\ref{sp_zeroT}) are explicitly written in the symmetric case, for $\rho=1$, where the replica method safely applies. They immediately recall those obtained by the cavity method, also implying:
\begin{equation}
    N^{*}(z)=\text{max}\left[0, \frac{K-\mu \langle N \rangle +z \sigma \sqrt{q_0}}{1-\rho \sigma^2 \chi} \right] \ ,
\end{equation}
where $\rho$ stands for the generic correlation between the interaction matrix couplings.

\vspace{0.3cm}

\subsection{Conclusions and Perspectives}

We shall now discuss a few exciting research directions that could pave the way not only for a better theoretical understanding of large ecosystems -- interpreted through the lens of disordered systems -- but also for a more immediate detection of empirical macroecological patterns and quantitative contributions to biomedical research.

\subsubsection{Dynamical Mean-Field Theory approach: asymmetric interactions and spatial dependence}

Non-reciprocal interactions break detailed balance. When the dynamics are no longer governed by conservative forces, defining a free energy and minimizing its Hessian on a proper subsector becomes unfeasible. However, asymmetric interactions can give rise to fascinating phenomena, including an exponential number of attractors, limit cycles, and chaotic behavior.

We now revisit the dynamical equations for the Generalized Lotka-Volterra model with arbitrary correlation $\rho$ between the interaction coefficients $A_{ij}$:
\begin{equation}
\frac{d N_i}{dt}=  N_i \left[(1-N_i) -\sum_{j,(j\neq i)} A_{ij}N_j \right] +\eta_i(t)+\lambda_i \ .
\end{equation}
In the large system size limit, for $S \rightarrow \infty$, this many-body problem can be conveniently rewritten in terms of a single-variable self-consistent stochastic process, also referred to as Dynamical Mean-Field Theory (DMFT):
\begin{equation}
\frac{dN(t)}{dt}=N(t) \left[ 1- N(t) -\mu h(t) -\sigma \zeta(t)+\rho \sigma^2\int_{0}^{t} dt' R(t,t') \; N(t) +H_\text{ext}(t) \right]
\end{equation}
where $h(t)= \mathbb{E}[N(t)]$ represents the mean abundance evolving in time, $C(t,t')=\mathbb{E}[N(t)N(t')]$ the correlation function, and $R(t,t')=\mathbb{E}\left[ \left.\frac{\delta N(t)}{\delta H_\text{ext}(t')} \right \vert_{H_\text{ext}=0} \right]$ the response function, thus computed w.r.t an external field $H_\text{ext}$. The notation $\mathbb{E}[\cdot]$ denotes the average on all different sources of randomness: random couplings, initial conditions, and demographic noise. 
The derivation is similar to that to obtain the Langevin equation from Newtonian dynamics. 
However, an additional layer of complexity arises here because the statistical properties of the effective thermal bath -- accounting for the interactions with the rest of the system -- must be determined self-consistently.

This DMFT formalism has been widely applied to a variety of interdisciplinary contexts, starting from strongly correlated electrons, to the analysis of rheological properties of amorphous systems under shear deformations, up to active matter. 
Moreover, this approach allows us to go beyond a purely well-mixed approximation and model a multi-species \emph{metacommunity} in the presence of external noise and random interactions \cite{garcia_lorenzana_interactions_2024}. 
In the simultaneous limits of an infinite number of species and spatial patches, the phase diagram can be determined analytically. For sufficiently large demographic fluctuations, the transition between an active phase -- where most species persist -- and an inactive phase is continuous, falling within the universality class of Directed Percolation. However, at lower values of demographic noise, the scenario changes dramatically. In this regime, the transition becomes discontinuous, unveiling novel features as well as the pivotal role of random interactions.

The effect of migration between distinct patches in
the strong heterogeneity regime was also investigated in the presence of non-symmetric interactions and zero noise \cite{roy2020, pearce2020stabilization}. In the latter, chaotic behaviors associated with long-lived persistent fluctuations can take place.

\subsubsection{Heavy-tailed distributions }

The RS solution discussed thus far predicts a truncated Gaussian distribution for the typical species abundances, which is known to be not the kind of distribution observed in Nature \cite{callaghan_unveiling_2023}. 
Nearly all the continuous statistical distributions (normal, exponential, gamma) have light tails. However, when talking about ecological collapses and external perturbations, such as the distribution of the largest fires within a limited region \cite{moritz1997}, the rate of population spread \cite{clark2001} or even hurricanes \cite{katz2002}, heavy-tailed distributions seem to be the right answer. 

Moreover, one might wonder whether the phase transitions obtained in high-dimensional phase diagrams (multiple equilibria, Gardner, etc.) remain true beyond the mean-field approximation, although
there is by now remarkable experimental
evidence of multiple alternative stable states.
In light of this, scaling and crossover theory might account for the long-range fluctuations that the mean-field approximation ignores.

%The search for universal laws in ecology typically leads to assuming a form of power-law shape (or Pareto distribution).

Within the framework of disordered systems, accurately representing more realistic distributions requires incorporating two additional key ingredients: asymmetric interactions and multiple fixed-point scenarios. Indeed, asymmetric interactions between species can result in chaotic dynamics, in which a sharp timescale separation between many rare species -- responsible for power-law trends -- and a few abundant ones, associated with sudden turnovers can be detected \cite{Kessler2015, xue2018, perkins2019}.
It is quite acknowledged that such heavy tails arise as an emergent properties of large ecological communities, the result of the interplay between interactions and stochasticity \cite{Grilli2021}. 
How a large number of species can coexist in those complex communities and what is the mechanism according to which they are mostly dominated by rare species is still not completely understood.
Over the years several efforts have been done to recover such distributions from simple statistical or mechanistic models:
\begin{itemize}
    \item {Heavy-tailed distributions can be obtained within an individual-based model (IBM) with a maximum number of individuals \cite{sole2002}, thus counterbalancing immigration of new species with the interactions between species. In other words, the IBM are stochastic cellular automatons in which immigration, growth rates, and extinction events can happen at every time step with the rates. All individuals of a species are assumed to be equivalent (same growth rates, same self-interaction strengths, same immigration parameters). Then, species abundances evolve stochastically. \cite{heyvaert2017, sole2002} showed separately that depending on the immigration rate, both power law and lognormal abundance distributions can be observed. By using experimental data of plankton and microbial communities, it has been found out that a log-normal distribution,
rather than a power law, fits best the distribution of abundances
of the most abundant species.}

\item{Alternatively, \cite{Grilli2021} proposes a stochastic GLV model with a global maximal capacity -- meaning that communities are limited by finite resources and limited space -- taking both the immigration and the degree of connectivity into account as relevant control parameters. At high connectivity, the abundance distribution allows for the emergence of non-trivial power-law behaviors.}

The abundance distribution of microbial communities is typically a lognormal.
In IBMs, broad distributions are the result of the system being self-organized at the edge of stability \cite{sole2002}. Immigration increases the number of species, whereas interactions between species cause individual extinction. Both the IBM and simplistic models with maximal capacity give rise, in different ways, to heavy-tailed rank abundance distributions. Notably, in the presence of logistic equations, a uniform distribution of the
self-interaction leads to a power-law abundance distribution. In contrast, lognormal distributions of the growth rate and the self-interaction term give rise to lognormal trends.

\end{itemize}

\subsubsection{Data-driven approaches}

The growing availability of observational datasets, particularly for microbial communities, allows increasingly detailed characterizations of local communities. 
Recent studies indicate that samples of the same community type, collected from various spatial or temporal locations, hosts, etc. frequently exhibit shared statistical properties \cite{ser-giacomi_ubiquitous_2018, grilli_macroecological_2020, gao_unifying_2024}. These findings support the application of ensemble averages as representative markers of distinct community states, a concept fundamental to statistical physics methodologies \cite{pasqualini_microbiomes_2024}.

%Differences across samples, environments, and community types (e.g., from various biomes or healthy versus diseased hosts) can be quantified, offering new ways to link ecological processes -- summarizing biological knowledge of species -- to functional differences at the community level. Recent studies suggest that multiple samples of the same community type, taken from different locations, hosts, or time points, often share common statistical features \cite{ser-giacomi_ubiquitous_2018,grilli_macroecological_2020, gao_unifying_2024}. This supports the use of ensemble averages as characteristic indicators of distinct community states, a concept central to statistical physics approaches \cite{pasqualini_microbiomes_2024}
%Identifying models that best describe community assembly requires complementing traditional abundance data with additional ecological information. 
However, comparing model predictions with empirical abundance distributions as a criterion for model selection may lead to several challenges.
Three notable approaches are usually employed in this regard.
First, longitudinal datasets reveal temporal changes in community composition, enabling more precise comparisons with theoretical models. Although deterministic and stochastic models may predict similar species abundance distributions, they differ in the timescales of fluctuations and resulting abundance trajectories. A key challenge requires distinguishing measurement errors, noise, and chaotic dynamics. New non-parametric time-series analysis methods are addressing these limitations \cite{rogers_chaos_2022}. Moreover, time-series properties like critical slowdown can signal approaching tipping points, potentially leading to catastrophic shifts in ecosystem states \cite{dakos_ecological_2022}. Detecting such transitions is crucial for ecological forecasting and decision-making, where statistical physics methods can play a fundamental role.

Second, incorporating ecological functions -- such as productivity, metabolic activity, or trophic structures -- into community descriptions can enhance the theoretical accuracy. The definition of \emph{functional groups}, accounting for the biological knowledge of how species with a similar role contributes to ecosystem functions, often reduces variability \cite{louca_function_2018,goldford_emergent_2018}. This multi-layer approach enables better model selection and motivates new theoretical directions. 

Third, some microbial communities exhibit ergodicity breaking, thus resulting in multiple attractor states \cite{hong2025, wright2021, lopes_cooperative_2024}. In light of this, since multistability reflects distinct community properties \cite{wright_inhibitory_2016, estrela_functional_2022}, there is hope that the inferred information from data could improve predictions on species compositions and emerging patterns, also for classifying specific ecosystem services and distinct diseases in microbial dynamics \cite{amor_smooth_2023, pasqualini_microbiomes_2024}.

\section*{Acknowledgements}

The author thanks Abishek Dhar, Joachim Krug, Satya N. Majumdar, Alberto Rosso, and Grégory Schehr for the organization of Les Houches Summer School 2024, \emph{Theory of Large Deviations and Applications}, where these lectures were delivered.
The author also acknowledges Yizhou Liu, Nirbhay Patil, and Frédéric van Wijland for a careful reading of the manuscript.

Explicit permission to reuse part of the material, specifically in Sections 3.2, 3.4, and 3.6, has been granted under APS license number RNP/25/JAN/087602.

\paragraph{Funding information}
A.A. acknowledges the support received
from the Agence Nationale de la Recherche (ANR) of the
French government, under the grant ANR-23-CE30-0012-
01 ("SIDECAR" project).

\vspace{0.75cm}

\bibliographystyle{apsr}
\bibliography{Notes-biblio}

\end{document}